# Very Low-Rate Variable-Length Channel Quantization for Minimum Outage Probability

Erdem Koyuncu and Hamid Jafarkhani

Center for Pervasive Communications and Computing

University of California, Irvine, Irvine CA 92697

## Abstract

We identify a practical vector quantizer design problem where any fixed-length quantizer (FLQ) yields non-zero distortion at any finite rate, while there is a variable-length quantizer (VLQ) that can achieve zero distortion with arbitrarily low rate. The problem arises in a $t \times 1$ multiple-antenna fading channel where we would like to minimize the channel outage probability by employing beamforming via quantized channel state information at the transmitter (CSIT). It is well-known that in such a scenario, finite-rate FLQs cannot achieve the full-CSIT (zero distortion) outage performance. We construct VLQs that can achieve the full-CSIT performance with finite rate. In particular, with $P$ denoting the power constraint of the transmitter, we show that the necessary and sufficient VLQ rate that guarantees the full-CSIT performance is $\Theta(1/P)$. We also discuss several extensions (e.g. to precoding) of this result.

## I. Introduction

High-resolution quantization theory says that for continuous sources with squared-error-like distortion measures, the distortion-rate functions for fixed-length quantizers (FLQs) and variable-length quantizers (VLQs) exhibit the same asymptotic behavior $\Theta(2^{-2R})$,[1] where the quantization rate $R$ is measured in bits per source sample [1]. This similarity in terms of the asymptotic behavior may not hold however for other distortion measures [2, Section VII]. In fact, one can easily construct distortion measures for which every finite-rate FLQ has non-zero distortion, while there is a VLQ that achieves zero distortion with finite rate. We first give an example for a discrete source and then extend it to the continuous case.

**Example 1.** *Consider a discrete source over an infinite alphabet $\{x_n : n \in \mathbb{N}\}$ with the Hamming distortion measure $h(x, \hat{x}) = \mathbf{1}(x \neq \hat{x})$ and probabilities $p_n \triangleq \mathtt{P}(x_n), n \in \mathbb{N}$ that satisfy $\sum_{n \in \mathbb{N}} p_n = 1$, $p_n > 0$, and $p_n \in O(\frac{1}{n^2})$. Any finite-rate FLQ then clearly has non-zero distortion. On the other hand, let $\mathtt{Q}$ be an infinite-level VLQ with $\mathtt{Q}(x_n) = x_n$ and codeword lengths $\ell_n \triangleq \lceil 2\log_2(n+1) + 1 \rceil$ (It is straightforward to show that $\sum_{n \in \mathbb{N}} 2^{-\ell_n} \leq 1$, and therefore, the code is prefix-free). Since $\ell_n \in \Theta(\log n)$ and $p_n \in O(\frac{1}{n^2})$, the quantization rate $\sum_{n \in \mathbb{N}} \ell_n p_n$ is finite.* ∎

The following is a continuous analogue of the discrete scenario in Example 1.

**Example 2.** *Let $p_n$ and $x_n$ be as defined in Example 1. For scalar quantization of the uniform distribution on $[0, 1]$, we construct a binary distortion measure $d_0(x, \hat{x}) :*

This work was supported in part by the NSF Award CCF-1218771.

[1]Let $f(x) \in O(g(x))$ if for sufficiently large $x$, $f(x) \leq ag(x)$ for some $a > 0$; $f(x) \in o(g(x))$ if for sufficiently large $x$, $f(x) \leq \epsilon g(x), \forall \epsilon > 0$; and $f(x) \in \Theta(g(x))$ if $f(x) \in O(g(x))$ and $f(x) \notin o(g(x))$.

$[0,1]^2 \to \{0,1\}$ *in the following manner: Let* $\mathcal{X}_n$, $n \in \mathbb{N}$ *be an arbitrary sequence of mutually disjoint subsets of* $[0,1]$ *with* $\bigcup_{n \in \mathbb{N}} \mathcal{X}_n = [0,1]$, $\mu(\mathcal{X}_n) = p_n$, *where* $\mu(\cdot)$ *denotes the Lebesgue measure. We set* $d_0(x, \hat{x}) = 0$ *whenever we have* $x \in \mathcal{X}_n$ *and* $\hat{x} = x_n$ *for some* $n \in \mathbb{N}$. *Otherwise, we set* $d_0(x, \hat{x}) = 1$. *Similarly, any finite-rate FLQ has positive distortion, while there is a finite-rate VLQ with zero distortion.* ∎

Example 2 shows that even for continuous sources, FLQ and VLQ distortion-rate functions may behave very differently. Given enough freedom, one can craft other distortion measures with the same (or perhaps even more "dramatic") consequences. Crafting, however, often results in an artifical distortion measure such as the one in Example 2. It is not clear whether such measures play any role in practical problems and deserve a thorough investigation. One goal of this paper is to demonstrate that in fact, they do appear in the context of multiple-antenna communication systems.

In multiple-antenna systems, the availability of channel state information (CSI) at the transmitter and/or the receiver can greatly improve the quality of communications. Typically, the receiver can acquire the CSI via training sequences from the transmitter. Obtaining CSI at the transmitter (CSIT) is however more difficult and generally requires feedback from the receiver. In practice, the feedback link has a finite bandwidth, which means that only a finite number of feedback bits per channel state can be utilized for feedback. One way to model such a limited feedback scenario is to formulate it as a source coding problem. The core element of such a formulation is a (channel) quantizer that specifies (i) for each channel state, the finite sequence of feedback bits to be fed back by the receiver; and (ii) for each such sequence, the codeword (e.g. a beamforming vector) to be employed by the transmitter.

A great deal of scholarly work has been done on the design and performance analysis of channel quantizers for multiple antenna systems; a comprehensive overview can be found in [3]. In particular, beamforming with limited feedback has been extensively studied with different approaches based on Grassmannian line packings [4], [5], vector quantizer design algorithms [6]–[8], high resolution approximations [9], random vector quantizers [10], and several other techniques [11]–[13].

Most of the previous work on finite-rate CSI feedback has been based on FLQs, in which the number of feedback bits per channel state is a fixed nonnegative integer. In general, different binary codewords of different lengths can be fed back for different channel states, resulting in a VLQ structure.

In this work, we consider the VLQ design problem for a multiple-input single-output (MISO) system with $t$ transmitter antennas and a short term power constraint $P$ at the transmitter. We assume a slowly fading channel model in which the channel realizations vary independently from one fading block to another while within each block they remain constant. We also assume that the receiver has full-CSI, while the transmitter has only partial CSI provided by the receiver via error-free and delay-free feedback channels. The partial CSI is in the form of quantized instantaneous CSI provided by a VLQ. Our performance measure is the outage probability.

Outage-optimal FLQs for MISO channels have been well-investigated [7], [11], and it is known that finite-rate FLQs cannot achieve the full-CSIT performance. We give an explicit construction of a VLQ that achieves the full-CSIT performance with rate $\mathsf{C}_0(\frac{1}{P} + \frac{1}{P^t})$, where $\mathsf{C}_0$ is a constant that is independent of $P$. Note that $\mathsf{C}_0(\frac{1}{P} + \frac{1}{P^t}) \to 0$ in the high signal-to-noise ratio (SNR) regime $P \to \infty$. We also show that such a decay $\Theta(\frac{1}{P})$ of the required feedback rate is asymptotically the best possible up to constant factors: For sufficiently high $P$, the performance of any VLQ with rate $o(\frac{1}{P})$ is strictly worse than the full-CSIT performance.

The rest of this paper is organized as follows: In Section II, we give a formal description of the system model, the outage probability performance measure, and the variable-length channel quantizers. In Section III, we state and prove our main results. Finally, in Section IV, we draw our main conclusions and discuss some extensions. A technical proof is provided in the appendix.

## II. Preliminaries

### A. System model

Consider a $t \times 1$ MISO system. Denote the channel from transmitter antenna $i$ to the receiver antenna by $h_i$, and let $\mathbf{h} = [\ h_1\ \cdots\ h_t\ ]^T \in \mathbb{C}^{t \times 1}$ represent the entire channel state. We assume that the entries of $\mathbf{h}$ are independent and identically distributed as $\mathtt{CN}(0,1)$. The transmitted symbol $\mathbf{s} \in \mathbb{C}^{t \times 1}$ and the received symbol $y \in \mathbb{C}$ have the input-output relationship $y = \mathbf{s}^T \sqrt{P} \mathbf{h} + \eta$, where the noise term $\eta \sim \mathtt{CN}(0,1)$ is independent of $\mathbf{h}$. We require $\mathtt{E}[\|\mathbf{s}\|^2] \leq 1$ as a result of the short-term power constraint of the transmitter.

### B. The outage probability performance measure

For a fixed $\mathbf{h}$, suppose that input symbol $\mathbf{s}$ is chosen as $\mathbf{s} \sim \mathtt{CN}(\mathbf{0}, \mathbf{xx}^\dagger)$ with $\|\mathbf{x}\| \leq 1$ due to the short-term power constraint. This transmission strategy is known as beamforming (along the beamforming vector $\mathbf{x}$). The channel capacity under this strategy is $\log_2(1 + |\langle \mathbf{x}, \mathbf{h} \rangle|^2 P)$ bits per channel use. For a given target data transmission rate $\rho$, we say that an outage event occurs if $\log_2(1 + |\langle \mathbf{x}, \mathbf{h} \rangle|^2 P) < \rho$.

When $\mathbf{h}$ is random, the transmitter can choose a different beamforming vector for different $\mathbf{h}$. In this case, we define the outage probability as the fraction of channel realizations for which outage events occur. Formally, consider an arbitrary (measurable) mapping $\mathtt{M} : \mathbb{C}^t \to \mathcal{X}_{\mathtt{BF}}$, where $\mathcal{X}_{\mathtt{BF}} \triangleq \{\mathbf{x} \in \mathbb{C}^{t \times 1} : \|\mathbf{x}\| = 1\}$ is the set of all feasible beamforming vectors (vectors with norm less than 1 can be shown to be suboptimal). Then, the outage probability with mapping $\mathtt{M}$ can be expressed as

$$\mathtt{OUT}(\mathtt{M}) \triangleq \mathtt{P}(|\langle \mathtt{M}(\mathbf{h}), \mathbf{h} \rangle|^2 < \alpha) \text{ with } \alpha = (2^\rho - 1)/P. \tag{1}$$

In the extreme case where the transmitter knows $\mathbf{h}$ perfectly (the full-CSIT case), it can choose an optimal beamforming vector, say $\mathtt{Full}(\mathbf{h})$ for a given $\mathbf{h}$. In such a scenario, we have $|\langle \mathtt{Full}(\mathbf{h}), \mathbf{h} \rangle| \leq \|\mathtt{Full}(\mathbf{h})\| \|\mathbf{h}\| \leq \|\mathbf{h}\|$, and the upper bound is achievable by choosing $\mathtt{Full}(\mathbf{h}) = \mathbf{h}/\|\mathbf{h}\|$. This gives us $\mathtt{OUT}(\mathtt{Full}) \triangleq \mathtt{P}(\|\mathbf{h}\|^2 < \alpha)$.

At the other extreme, the transmitter may have "no idea" about $\mathbf{h}$, in which case we say that we have a no-CSIT system. In such a scenario, the transmitter has to use a single beamforming vector, say $\mathbf{x}_0 \in \mathcal{X}_{\mathtt{BF}}$, for all the channel states. This gives the outage probability $\mathtt{P}(|\langle \mathbf{x}_0, \mathbf{h} \rangle|^2 < \alpha)$, which can be shown to be independent of $\mathbf{x}_0$. We may therefore set $\mathtt{Open}(\mathbf{h}) \triangleq \mathbf{e}_1, \forall \mathbf{h}$ without loss of optimality, where $\mathbf{e}_1 = [\ 1\ 0\ \cdots\ 0\ ]^\dagger$. We have $\mathtt{OUT}(\mathtt{Open}) = \mathtt{P}(|h_1|^2 < \alpha)$.

We now consider the case where the transmitter has partial CSI via feedback from the receiver. Using a source coding formulation, such a partial CSI system can be described by a channel quantizer as we explain in what follows.

*C. The channel quantizer*

Let $\mathcal{I} \subset \mathbb{N}$ be an index set. We use the notations $\{a_n\}_\mathcal{I}$ and $\{a_n : n \in \mathcal{I}\}$ interchangeably to represent a set whose elements are the real numbers $a_n$, $n \in \mathcal{I}$. For $\mathcal{I} = \mathbb{N}$, we ignore the subscript and write $\{a_n\}$ instead of $\{a_n\}_\mathbb{N}$. Similar definitions hold for sets of vectors, collection of sets, etc.

For a given index set $\mathcal{I}$, let $\{\mathbf{x}_n\}_\mathcal{I}$ be a set of quantized beamforming vectors with $\{\mathbf{x}_n\}_\mathcal{I} \subset \mathcal{X}_{\texttt{BF}}$. Also, let $\{\mathcal{E}_n\}_\mathcal{I}$ with $\mathcal{E}_n \subset \mathbb{C}^t$, $\forall n \in \mathbb{N}$ be a collection of mutually disjoint measurable subsets of $\mathbb{C}^t$ with $\bigcup_{n \in \mathcal{I}} \mathcal{E}_n = \mathbb{C}^t$. Finally, let $\{\mathbf{b}_n\}_\mathcal{I}$ be a collection of feedback binary codewords with $\{\mathbf{b}_n\}_\mathcal{I} \subset \{0,1\}^\star$, where $\{0,1\}^\star \triangleq \{\epsilon, 0, 1, 00, 01, \ldots\}$ is the set of all binary codewords including the empty codeword $\epsilon$. We assume $\mathbf{b}_m \neq \mathbf{b}_n$ whenever $m \neq n$. We call the collection of triples

$$\texttt{Q} := \{(\mathbf{x}_n, \mathcal{E}_n, \mathbf{b}_n)\}_\mathcal{I} \qquad (2)$$

a quantizer $\texttt{Q}$ for the beamforming strategy. We call $\texttt{Q}$ an infinite-level quantizer if $\mathcal{I}$ is an infinite set. Otherwise, we call $\texttt{Q}$ an $|\mathcal{I}|$-level quantizer.

The quantizer definition in (2) immediately induces a feedback transmission scheme that operates in the following manner: For a fixed channel state $\mathbf{h}$, the receiver feeds back the binary codeword $\mathbf{b}_n$, where the index $n$ here satisfies $\mathbf{h} \in \mathcal{E}_n$. Such an index $n$ always exists and is unique as $\mathcal{E}_n$, $n \in \mathbb{N}$ is a disjoint covering of $\mathbb{C}^t$. The transmitter recovers the index $n$ and uses the corresponding beamforming vector $\mathbf{x}_n$. The recovery of $n$ by the transmitter is always possible since $\mathbf{b}_n$s are distinct. We write $\texttt{Q}(\mathbf{h}) = \mathbf{x}_n$ whenever $\mathbf{h} \in \mathcal{E}_n$ to emphasize the quantization operation. We call the set $\{\mathbf{x}_n\}_\mathcal{I}$ the quantizer (or beamforming) codebook.

For any $\mathbf{b} \in \{0,1\}^\star$, let $\texttt{L}(\mathbf{b})$ denote the "length" of $\mathbf{b}$. We allow $\texttt{L}(\mathbf{b}) = \infty$ if $\mathbf{b}$ is not of finite length (For example, $\texttt{L}(\epsilon) = 0$, $\texttt{L}(01) = 2$, $\texttt{L}(1010\cdots) = \infty$). A quantizer $\texttt{Q}$ is called an FLQ if $\texttt{L}(b_m) = \texttt{L}(b_n)$, $\forall m, n \in \mathcal{I}$. Otherwise, we call $\texttt{Q}$ a VLQ.

A quantizer $\texttt{Q}$ is thus a mapping $\mathbb{C}^t \to \{\mathbf{x}_n\}_\mathcal{I}$ supplied with a feedback binary codeword $\mathbf{b}_n$ for each $\mathbf{x}_n$. Treated solely as a mapping, it is a special case of the mapping $\texttt{M} : \mathbb{C}^t \to \mathcal{X}_{\texttt{BF}}$ discussed in the previous section with the requirement of a countable image $\{\mathbf{x}_n\}_\mathcal{I}$. According to (1), we can therefore calculate the outage probability with $\texttt{Q}$ as $\texttt{OUT}(\texttt{Q}) = \texttt{P}(|\langle \texttt{Q}(\mathbf{h}), \mathbf{h}\rangle|^2 < \alpha)$, which does not depend on $\mathbf{b}_n$. As a result, we do not specify/mention $\mathbf{b}_n$ when we would like to talk only about the outage performance of a quantizer and write $\{\mathbf{x}_n, \mathcal{E}_n, \cdot\}_\mathcal{I}$ instead. The binary codewords $\mathbf{b}_n$ however determine the rate $\texttt{R}(\texttt{Q})$ of the quantizer $\texttt{Q}$ via $\texttt{R}(\texttt{Q}) \triangleq \sum_{n \in \mathcal{I}} \texttt{P}(\mathbf{h} \in \mathcal{E}_n) \texttt{L}(\mathbf{b}_n)$.

Clearly, we measure the quality of the quantizer by the outage probability it provides. We can also define the distortion measure $d(\mathbf{h}, \mathbf{x}) = \mathbf{1}(|\langle \mathbf{x}, \mathbf{h}\rangle|^2 < \alpha \vee \|\mathbf{h}\|^2 < \alpha) = \mathbf{1}(|\langle \mathbf{x}, \mathbf{h}\rangle|^2 < \alpha) - \mathbf{1}(\|\mathbf{h}\|^2 < \alpha)$ that measures the quality of reproduction of the source/channel sample $\mathbf{h}$ by $\mathbf{x}$. For a given quantizer $\texttt{Q}$, the expected distortion $\texttt{E}[d(\mathbf{h}, \texttt{Q}(\mathbf{h}))]$ is nothing but the quantity $\texttt{OUT}(\texttt{Q}) - \texttt{OUT}(\texttt{Full})$. Therefore, minimizing the expected distortion is equivalent to minimizing the outage probability with $\texttt{Q}$.

Note that $d(\mathbf{h}, \mathbf{x}) \in \{0, 1\}$ for any $\mathbf{h}$ and $\mathbf{x}$, and thus $d(\mathbf{h}, \mathbf{x})$ is a binary-distortion function for a continous source. In this respect, it resembles the function $d_0(x, \hat{x})$ in Example 2. Also similarly, for any given reproduction level $\mathbf{x}$, there is a set of source samples, say $\mathcal{H}_\mathbf{x}$, that has positive probability and satisfies $\forall \mathbf{h} \in \mathcal{H}_\mathbf{x}, d(\mathbf{h}, \mathbf{x}) = 0$. We thus expect VLQs to signficantly outperform FLQs for the distortion measure $d(\mathbf{h}, \mathbf{x})$. On the other hand, it is not clear how to design a good VLQ for $d(\mathbf{h}, \mathbf{x})$, as unlike Example 2, the question of how to design the quantizer codebook and the encoding regions is non-trivial.

## III. Finite-rate VLQs that achieve minimum outage

We start with the design of the encoding regions for a given codebook. To gain initial insight on the problem, we first review the standard encoding rule used for FLQs and discuss why it will not work in the case of VLQs.

### A. The standard encoding rule

Let $\mathcal{B} = \{\mathbf{x}_0, \ldots, \mathbf{x}_{N-1}\}$ be a finite-cardinality codebook. A standard practice (see e.g. [7], [11]) is to work with the quantizer $\bar{\mathsf{Q}}_\mathcal{B}(\mathbf{h}) \triangleq \arg\max_{\mathbf{x} \in \mathcal{B}} |\langle \mathbf{x}, \mathbf{h} \rangle|^2$, which chooses the beamforming vector (with ties broken artbirarily) in $\mathcal{B}$ that is "closest" to $\mathbf{h}$. In fact, it can be shown that $\bar{\mathsf{Q}}_\mathcal{B}$ is an optimal quantizer for the codebook $\mathcal{B}$ in the sense that for any quantizer $\mathsf{Q}_\mathcal{B} : \mathbb{C}^t \to \mathcal{B}$, we have $\mathtt{OUT}(\bar{\mathsf{Q}}_\mathcal{B}) \leq \mathtt{OUT}(\mathsf{Q}_\mathcal{B})$.

One way to design a VLQ might be to keep the standard encoding rule but instead use a variable-length code instead of a fixed-length code. There are two problems with this approach. The first problem, which is of a rather technical nature, is that the standard encoding rule is not well-defined for infinite-level quantizers as a maximizer may not exist for countably infinite codebooks. The second and more important issue is that even for a finite cardinality codebook, this rule is quite ill-suited for variable-length quantization as we shall discuss in the following.

It is well known that the beamforming vectors in a well-designed codebook should be "evenly distributed" on $\mathcal{X}_{\mathtt{BF}}$ (A formal treatment of this argument gives rise to e.g. Grassmannian codebooks [4]). For such a well-designed codebook $\mathcal{B} = \{\mathbf{x}_0, \ldots, \mathbf{x}_{N-1}\}$ and an index $i \in \{0, \ldots, N-1\}$, the standard encoder picks $\mathbf{x}_i$ if $\forall n \in \{0, \ldots, N-1\}, |\langle \mathbf{x}_i, \mathbf{h} \rangle| \geq |\langle \mathbf{x}_n, \mathbf{h} \rangle|$. Due to the even distribution of codevectors, this results in quantization cells with roughly equal probability $\frac{1}{N}$. In such a scenario, it can be shown that even the best variable-length code results in a VLQ rate of $\log_2 N$ (up to an additive constant). Hence, VLQs designed via the standard encoding rule cannot achieve the full-CSIT performance with finite rate since we need $N \to \infty$ (In fact, we can already design a rate-$\lceil \log_2 N \rceil$ FLQ that is optimal for $\mathcal{B}$; a VLQ with almost the same rate is superfluous). We thus first introduce an alternate encoding strategy.

### B. An alternate encoding rule

For any given beamforming vector $\mathbf{x} \in \mathcal{X}_{\mathtt{BF}}$, let $\mathcal{O}_\mathbf{x} = \{\mathbf{h} : |\langle \mathbf{x}, \mathbf{h} \rangle| < \sqrt{\alpha}\}$ denote the channel states for which using $\mathbf{x}$ results in outage. Also, let $\mathcal{O}_\mathbf{x}^c$ denote the complement of $\mathcal{O}_\mathbf{x}$. The simple but key observation is that the standard encoder is "excessively precise" as it always picks the (intuitively best) beamforming vector in $\{\mathbf{x}_0, \ldots, \mathbf{x}_{N-1}\}$ that is closest to $\mathbf{h}$. In fact, without loss of optimality in terms of the outage probability, for any $j \in \{0, \ldots, N-1\}$, the transmitter can use $\mathbf{x}_j$ whenever using $\mathbf{x}_j$ does not result in outage (i.e. whenever $\mathbf{h} \in \mathcal{O}_{\mathbf{x}_j}^c$). It can also use $\mathbf{x}_j$ whenever all the beamforming vectors in the codebook result in outage (i.e. whenever $\mathbf{h} \in \bigcap_{n=0}^{N-1} \mathcal{O}_{\mathbf{x}_n}$), as using any other beamforming vector in $\{\mathbf{x}_0, \ldots, \mathbf{x}_{N-1}\}$ would result in an inevitable outage anyway. In other words, for the set $\mathcal{O}_{\mathbf{x}_j} \cup \bigcap_{n=0}^{N-1} \mathcal{O}_{\mathbf{x}_n}$ of source samples, choosing $\mathbf{x}_j$ instead of the beamforming vector that is closest to $\mathbf{h}$ will not change the distortion. We exploit this property of the distortion measure to design a new encoding strategy that yields very low rates without sacrificing performance.

Formally, for a given arbitrary infinite beamforming codebook $\{\mathbf{x}_n\}$, we set

$$\mathcal{E}_0^\star \triangleq \mathcal{O}_{\mathbf{x}_0}^c \cup \bigcap_{n \in \mathbb{N}} \mathcal{O}_{\mathbf{x}_n}, \tag{3}$$

and use $\mathbf{x}_0$ as the beamforming vector whenever $\mathbf{h} \in \mathcal{E}_0^\star$. We have now allocated the part $\mathcal{E}_0^\star$ of the entire channel state space $\mathbb{C}^t$. In general, for any $n \in \mathbb{N} - \{0\}$, we set

$$\mathcal{E}_n^\star = \mathcal{O}_{\mathbf{x}_n}^c \cap \bigcap_{k=0}^{n-1} \mathcal{O}_{\mathbf{x}_k}, \tag{4}$$

and use $\mathbf{x}_n$ whenever $\mathbf{h} \in \mathcal{E}_n^\star$. For any $n \in \mathbb{N} - \{0\}$, by definition, $\mathcal{E}_n^\star$ consists of channel states for which using the beamforming vector $\mathbf{x}_n$ does not result in outage while using any of the preceding beamforming vectors $\mathbf{x}_0, \ldots, \mathbf{x}_{n-1}$ results in outage.

It follows immediately from the definitions that $\{\mathcal{E}_n^\star\}$ is a disjoint collection of measurable sets that cover $\mathbb{C}^t$. We may therefore define the infinite-level quantizer $\{\mathbf{x}_n, \mathcal{E}_n^\star, \cdot\}$. Let us now calculate the outage probability with this quantizer. Meanwhile, we show that it is also in fact an optimal quantizer for the codebook $\{\mathbf{x}_n\}$.

**Proposition 1.** *Let $\{\mathbf{x}_n\}$ be a given codebook. For any quantizer $\mathtt{Q} : \mathbb{C}^t \to \{\mathbf{x}_n\}$, we have $\mathtt{OUT}(\mathtt{Q}) \geq \mathsf{P}\left(\mathbf{h} \in \bigcap_{i \in \mathbb{N}} \mathcal{O}_{\mathbf{x}_i}\right)$. In particular, $\mathtt{OUT}(\{\mathbf{x}_n, \mathcal{E}_n^\star, \cdot\}) = \mathsf{P}\left(\mathbf{h} \in \bigcap_{i \in \mathbb{N}} \mathcal{O}_{\mathbf{x}_i}\right)$, and therefore $\{\mathbf{x}_n, \mathcal{E}_n^\star, \cdot\}$ is an optimal quantizer for the codebook $\{\mathbf{x}_n\}$.*

*Proof:* For any quantizer $\mathtt{Q} = \{\mathbf{x}_n, \mathcal{E}_n, \cdot\}$, the event $\mathbf{h} \in \bigcap_{n \in \mathbb{N}} \mathcal{O}_{\mathbf{x}_n}$ results in outage regardless of how $\{\mathcal{E}_n\}$ is chosen. This proves the lower bound. As for the quantizer $\{\mathbf{x}_n, \mathcal{E}_n^\star, \cdot\}$, by construction, an outage event happens if and only if $\mathbf{h} \in \mathcal{E}_0^\star = \mathcal{O}_{\mathbf{x}_0}^c \cup \bigcap_{n \in \mathbb{N}} \mathcal{O}_{\mathbf{x}_n}$, for which the transmitter uses the beamforming vector $\mathbf{x}_0$. Since $\mathbf{x}_0$ does not result in outage when $\mathbf{h} \in \mathcal{O}_{\mathbf{x}_0}^c$, we have the desired result. ■

For a finite-cardinality codebook $\{\mathbf{x}_n : n = 0, \ldots, N-1\}$, we replace the "$\mathbb{N}$" in the right-hand side of (3) by "$\{0, \ldots, N-1\}$," and follow the same steps (e.g. Proposition 1) to construct an optimal quantizer for codebook $\{\mathbf{x}_n : n = 0, \ldots, N-1\}$.

We now argue that with an appropriate choice for $\{\mathbf{x}_n\}$ and the feedback binary codewords $\{\mathbf{b}_n\}$ the quantizer $\mathtt{Q}_{\{\mathbf{x}_n\}}^\star$ can achieve $\mathtt{OUT}(\mathtt{Full})$ with finite rate. We first give a "proof by figures." For this purpose, the regions $\mathcal{O}_\mathbf{x} = \{\mathbf{h} : |\langle \mathbf{x}, \mathbf{h}\rangle| < \sqrt{\alpha}\}$ and $\mathcal{O}_\mathbf{x}^c$ are illustrated in Fig. 1a for some $\alpha > 1$. In the figure, the entire space $\mathbb{C}^t$ is represented by the interior of the outer "disk" $\|\mathbf{h}\| \leq \infty$ that is bounded by the "circle" $\|\mathbf{h}\| = \infty$. The inner disk represents the sphere $\|\mathbf{h}\|^2 \leq \alpha$. In fact, the probability that $\mathbf{h}$ remains in the interior of this disk is the full-CSIT outage probability $\mathtt{OUT}(\mathtt{Full})$. The beamforming vector $\mathbf{x}$ resides on the sphere $\|\mathbf{h}\| = 1$ (not shown). The lighter shaded region in the middle is $\mathcal{O}_\mathbf{x}$ and the remaining two darker shaded regions constitute $\mathcal{O}_\mathbf{x}^c$. Note that the regions $\mathcal{O}_\mathbf{x}$ and $\mathcal{O}_\mathbf{x}^c$ are in general separated by the two hyperplanes $\{\mathbf{h} : \langle \mathbf{x}, \mathbf{h}\rangle = \sqrt{\alpha}\}$ and $\{\mathbf{h} : \langle \mathbf{x}, \mathbf{h}\rangle = -\sqrt{\alpha}\}$ that are perpendicular to $\mathbf{x}$. The two parallel lines in the figure represent these hyperplanes.

Using the finite codebook versions of (3) and (4), the encoding regions given a quantizer codebook $\{\mathbf{x}_0, \ldots, \mathbf{x}_3\}$ will then be as shown as in Fig. 1b. In the figure, $\mathcal{E}_0^\star$ comprises of the interior of the hexagon formed at the center, and the two half planes on the left and right sides of the figure. By Proposition 1, the probability

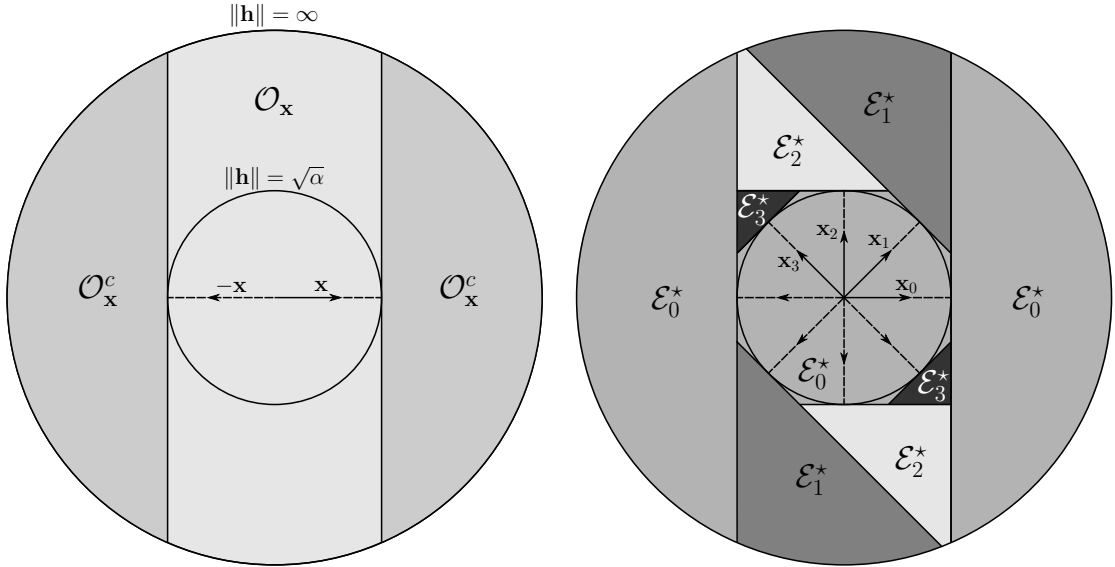

(a) Outage ($\mathcal{O}_\mathbf{x}$) and no-outage ($\mathcal{O}_\mathbf{x}^c$) regions of $\mathbf{x}$.   (b) The encoding regions of a 4-level quantizer.

Fig. 1: An illustration of the new encoding rule.

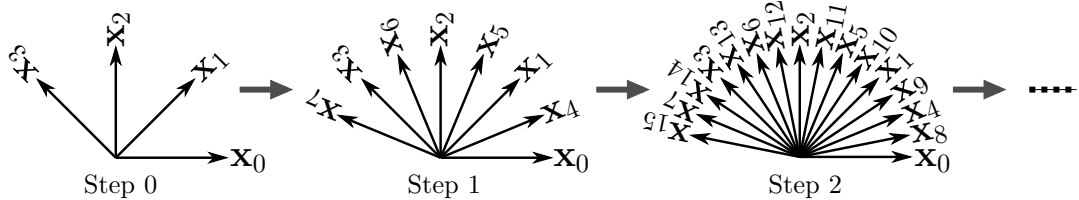

Fig. 2: A layered codebook structure.

that $\mathbf{h}$ remains in the interior of the hexagon is the outage probability with the quantizer $\{\mathbf{x}_n, \mathcal{E}_n^\star, \cdot\}_{0 \leq n \leq 3}$. It is greater than the full-CSIT outage probability as the hexagon cannot "completely cover" the inner circle. Now, as shown in Fig. 2, at Step 0, we start with the codebook $\{\mathbf{x}_0, \ldots, \mathbf{x}_3\}$ in Fig. 1b, and at Step $\ell$, we add $2^{\ell+1}$ new beamforming vectors in between the ones we had in Step $\ell - 1$. Repeating this process indefinitely gives us an infinite codebook $\{\mathbf{x}_n\}$ with a layered structure.

If we were to draw the codebook $\{\mathbf{x}_n\}$ as we drew $\{\mathbf{x}_0, \ldots, \mathbf{x}_3\}$ we would observe that now $\bigcap_{n \in \mathbb{N}} \mathcal{O}_\mathbf{x}$ coincides with $\|\mathbf{h}\| < \alpha$ (up to a null set). This means that $\{\mathbf{x}_n, \mathcal{E}_n^\star, \cdot\}$ can achieve $\mathtt{OUT}(\mathtt{Full})$. We now specify the feedback binary codewords for each quantization cell. Let $\mathtt{b}_0^\star = \epsilon$, $\mathtt{b}_1^\star = \mathtt{0}$, $\mathtt{b}_2^\star = \mathtt{1}$, $\mathtt{b}_3^\star = \mathtt{00}$, $\mathtt{b}_4^\star = \mathtt{01}$, and sequentially so on for all the feedback binary codewords in $\{0, 1\}^\star$. We have $\mathtt{L}(\mathtt{b}_n^\star) = \lfloor \log_2(n+1) \rfloor$, and we consider the rate of the quantizer $\{\mathbf{x}_n, \mathcal{E}_n^\star, \mathtt{b}_n^\star\}$. From our figure for the encoding regions on $\{\mathbf{x}_n\}$, we would also observe that the probabilities $\mathtt{P}(\mathbf{h} \in \mathcal{E}_n^\star)$ decay rather fast (If we take our two-dimensional "paper world" to be exact, they decay as $\frac{1}{n^2}$). We show that they in fact decay fast enough so that the quantization rate $\sum_{n \in \mathbb{N}} \mathtt{P}(\mathbf{h} \in \mathcal{E}_n^\star) \mathtt{L}(\mathtt{b}_n^\star)$ is finite. Our analysis reveals that the rate is bounded above by $\mathtt{C}_0(\frac{1}{P} + \frac{1}{P^t})$ for some constant $\mathtt{C}_0$. In the following, we formalize these claims. We first construct a codebook that has the layered structure in Fig. 2.

*C. A layered codebook*

For any $\ell \in \mathbb{N}$, let $\mathcal{S}_\ell = \{-1 + \frac{k}{2^{\ell+1}}, k = 1, \ldots, 2^{\ell+2} - 1\}$. For example, we have $\mathcal{S}_0 = \{-\frac{1}{2}, 0, \frac{1}{2}\}$, and $\mathcal{S}_1 = \{-\frac{3}{4}, -\frac{1}{2}, -\frac{1}{4}, 0, \frac{1}{4}, \frac{1}{2}, \frac{3}{4}\}$. For a given $\ell$, we construct a codebook $\mathcal{Y}_\ell$ of beamforming vectors by setting

$$\mathcal{Y}_\ell = \{\mathbf{y}/\|\mathbf{y}\| : \Re y_i \in \mathcal{S}_\ell, \Im y_i \in \mathcal{S}_\ell, i = 1, \ldots, t, \text{ and } 0 < \|\mathbf{y}\| \leq 1\},$$

where $\Re y_i$ and $\Im y_i$ denote the real and imaginary parts of $y_i$, respectively. For increasing $\ell$, $\mathcal{S}_\ell$ provides a fine quantization of the interval $[-1, 1]$. The corresponding $\mathcal{Y}_\ell$ is roughly a product quantizer codebook, and thus provides an increasingly finer quantization of $\mathcal{X}_{\text{BF}}$ as $\ell$ increases. In fact, the sequence of codebooks $\{\mathcal{Y}_\ell\}_{\ell \in \mathbb{N}}$ looks like the sequence of layered codebooks in Fig. 2, with the index "$\ell$" representing the layer. Formally, we have the following.

**Proposition 2.** *For any $\ell \in \mathbb{N}$, $\mathcal{Y}_\ell \subset \mathcal{Y}_{\ell+1}$, and $|\mathcal{Y}_\ell| \leq 2^{2t(\ell+2)}$. Also, for sufficiently large $\ell$, $\forall \overline{\mathbf{h}} \in \mathcal{X}_{\text{BF}}, \exists \mathbf{y} \in \mathcal{Y}_\ell, |\langle \mathbf{y}, \overline{\mathbf{h}} \rangle|^2 > 1 - \frac{2t}{2^\ell}.$*

*Proof:* The bound on $|\mathcal{Y}_\ell|$ and the fact that $\mathcal{Y}_\ell \subset \mathcal{Y}_{\ell+1}, \forall n \in \mathbb{N}$ follow immediately from the definitions. We now prove the last property.

Let $\epsilon = \frac{1}{2^{\ell+1}}$. We can then rewrite $\mathcal{S}_\ell = \{-1 + \epsilon k, k = 1, \ldots, \frac{2}{\epsilon} - 1\}$. For any real number $x \in [-1, 1]$, we consider the scalar quantizer

$$q(x) = \begin{cases} -1 + k\epsilon, & \exists k \in \{1, \ldots, \frac{1}{\epsilon} - 1\}, x \in [-1 + (k-1)\epsilon, -1 + k\epsilon) \\ 0, & x \in [-\epsilon, \epsilon] \\ -1 + (k-1)\epsilon, & \exists k \in \{\frac{1}{\epsilon} + 2, \ldots, \frac{2}{\epsilon}\}, x \in (-1 + (k-1)\epsilon, -1 + k\epsilon] \end{cases}. \quad (5)$$

We can observe that for any $x \in [-1, 1]$, $q(x)$ is well-defined with $q(x) \in \mathcal{S}_\ell$,

$$|q(x)| \leq |x|, \quad (6)$$

and

$$|q(x) - x| \leq \epsilon. \quad (7)$$

The last two properties (6) and (7) imply in particular that

$$|x| \leq |q(x)| + \epsilon, \quad (8)$$

by the reverse triangle inequality.

Now, for any given $\overline{\mathbf{h}} = [\overline{h}_1 \cdots \overline{h}_t]^T \in \mathcal{X}_{\text{BF}}$ as stated in the lemma, let $\mathbf{z} = [z_1 \cdots z_t]^T$ be given by

$$\Re z_i = q(\Re \overline{h}_i), \Im z_i = q(\Im \overline{h}_i), i = 1, \ldots, t. \quad (9)$$

We have $\Re z_i, \Im z_i \in \mathcal{S}_\ell$. Let us now verify that in fact $0 < \|\mathbf{z}\| \leq 1$. Suppose that $\|\mathbf{z}\| = 0$. Then, $\mathbf{z}$ is the all-zero vector, or equivalently $q(\Re \overline{h}_i) = q(\Im \overline{h}_i) = 0, i = 1, \ldots, t$ by the definition of $\mathbf{z}$. According to (8), we then have

$$\|\overline{\mathbf{h}}\|^2 = \sum_{i=1}^{t} \left(|\Re \overline{h}_i|^2 + |\Im \overline{h}_i|^2\right)$$

$$\leq \sum_{i=1}^{t} \left(|q(\Re \overline{h}_i) + \epsilon|^2 + |q(\Im \overline{h}_i) + \epsilon|^2\right)$$

$$= 2t\epsilon^2$$
$$= \frac{2t}{2^{2(\ell+1)}}$$
$$< 1,$$

where the last inequality holds for sufficiently large $\ell$. This contradicts the fact that $\|\overline{\mathbf{h}}\| = 1$. On the other hand, if $\|\mathbf{z}\| > 1$, according to (7) and (9), the inequalities

$$\begin{aligned}
1 &< \|\mathbf{z}\|^2 \\
&= \sum_{i=1}^{t}\left(|\Re z_i|^2 + |\Im z_i|^2\right) \\
&= \sum_{i=1}^{t}\left(|q(\Re \overline{h}_i)|^2 + |q(\Im \overline{h}_i)|^2\right) \\
&\leq \sum_{i=1}^{t}\left(|\Re \overline{h}_i|^2 + |\Im \overline{h}_i|^2\right) \\
&= \|\overline{\mathbf{h}}\|^2 \\
&= 1
\end{aligned}$$

lead to a contradiction. We have thus established $\Re z_i, \Im z_i \in \mathcal{S}_\ell$ with $0 < \|\mathbf{z}\| \leq 1$. We now show that $|\langle \mathbf{z}, \overline{\mathbf{h}}\rangle|^2 > 1 - \frac{2t}{2^\ell}$ so that the last property in the proposition holds with $\mathbf{y} = \mathbf{z}/\|\mathbf{z}\|$. For this purpose, let us first obtain a lower estimate for $\|\mathbf{z}\|^2$. By (8) and (9), we have $|\Re z_i| \geq |\Re \overline{h}_i| - \epsilon$, $|\Im z_i| \geq |\Im \overline{h}_i| - \epsilon$, $i = 1, \ldots, t$. Now, for any $i \in \{1, \ldots, t\}$, if $|\Re \overline{h}_i| \geq \epsilon$, we have

$$|\Re z_i|^2 \geq (|\Re \overline{h}_i| - \epsilon)^2 > |\Re \overline{h}_i|^2 - 2|\Re \overline{h}_i|\epsilon.$$

Otherwise, if $|\Re \overline{h}_i| < \epsilon$, $|\Re z_i|^2 = 0$ by the definition of $q(\cdot)$, and therefore,

$$|\Re z_i|^2 = 0 \geq |\Re \overline{h}_i|^2 - 2|\Re \overline{h}_i|\epsilon,$$

as the function $x^2 - 2x\epsilon$ is non-positive for all $x \in [0, 2\epsilon]$. Combining the two cases, the inequality

$$|\Re z_i|^2 \geq |\Re \overline{h}_i|^2 - 2|\Re \overline{h}_i|\epsilon$$

holds for any $i \in \{1, \ldots, t\}$. Noting that a similar set of inequalities holds for $|\Im z_i|^2$, $i = 1, \ldots, t$, we obtain

$$\begin{aligned}
\|\mathbf{z}\|^2 &= \sum_{i=1}^{t}\left(|\Re z_i|^2 + |\Im z_i|^2\right) \\
&> \sum_{i=1}^{t}\left(|\Re \overline{h}_i|^2 - 2|\Re \overline{h}_i|\epsilon + |\Im \overline{h}_i|^2 - 2|\Im \overline{h}_i|\epsilon\right) \\
&= 1 - 2\epsilon \sum_{i=1}^{t}\left(|\Re \overline{h}_i| + |\Im \overline{h}_i|\right).
\end{aligned}$$

Subject to $\|\overline{\mathbf{h}}\|^2 = 1$, we have $\sum_{i=1}^{t}(|\Re \overline{h}_i| + |\Im \overline{h}_i|) \leq \sqrt{2t}$ with equality if and only if $|\Re \overline{h}_i| = |\Im \overline{h}_i| = \frac{1}{\sqrt{2t}}$, $\forall i \in \{1, \ldots, t\}$. Therefore,

$$\|\mathbf{z}\|^2 > 1 - 2\sqrt{2t}\epsilon. \tag{10}$$

Moreover, according to (7) and (9), we have

$$\|\overline{\mathbf{h}} - \mathbf{z}\|^2 \leq 2t\epsilon^2. \tag{11}$$

We can now use the decomposition

$$\|\overline{\mathbf{h}} - \mathbf{z}\|^2 = (\overline{\mathbf{h}} - \mathbf{z})^\dagger (\overline{\mathbf{h}} - \mathbf{z})$$
$$= 1 + \|\mathbf{z}\|^2 - 2\Re\langle\overline{\mathbf{h}}, \mathbf{z}\rangle.$$

Isolating $\Re\langle\overline{\mathbf{h}}, \mathbf{z}\rangle$ and then using (10) and (11), we obtain

$$\Re\langle\overline{\mathbf{h}}, \mathbf{z}\rangle = \frac{1}{2}\left(1 + \|\mathbf{z}\|^2 - \|\overline{\mathbf{h}} - \mathbf{z}\|^2\right)$$
$$> \frac{1}{2}\left(1 + (1 - 2\sqrt{2t}\epsilon) - 2t\epsilon^2\right)$$
$$= 1 - \underbrace{\sqrt{2t}}_{\leq t}\epsilon - t\underbrace{\epsilon^2}_{<\epsilon}$$
$$> 1 - 2t\epsilon.$$

Therefore,

$$|\langle\overline{\mathbf{h}}, \mathbf{z}\rangle|^2 \geq (\Re\langle\overline{\mathbf{h}}, \mathbf{z}\rangle)^2 > (1 - 2t\epsilon)^2 > 1 - 4t\epsilon.$$

Letting $\mathbf{y} \triangleq \frac{\mathbf{z}}{\|\mathbf{z}\|}$, we have the claimed $\mathbf{y} \in \mathcal{Y}_\ell$ with $|\langle\mathbf{y}, \overline{\mathbf{h}}\rangle|^2 > 1 - 4t\epsilon = 1 - \frac{2t}{2^\ell}$. ∎

There is a quantizer codebook $\{\mathbf{y}_n\}_\mathbb{N}$ with the property that $\forall \ell \in \mathbb{N}$, $\bigcup_{n=0}^{|\mathcal{Y}_\ell|-1}\{\mathbf{y}_n\} = \mathcal{Y}_\ell$. In other words, the "first" $|\mathcal{Y}_0|$ elements $\mathbf{y}_0, \ldots, \mathbf{y}_{|\mathcal{Y}_0|-1}$ of $\{\mathbf{y}_n\}$ form the set $\mathcal{Y}_0$, the next $|\mathcal{Y}_1|$ elements of $\{\mathbf{y}_n\}$ form the set $\mathcal{Y}_1$, and so on. Such a set $\{\mathbf{y}_n\}$ always exists since by Proposition 2, $\mathcal{Y}_\ell \subset \mathcal{Y}_{\ell+1}$ and each $\mathcal{Y}_\ell$ has finite cardinality. In fact, it is also straightforward to construct $\{\mathbf{y}_n\}$ by setting $\{\mathbf{y}_0, \ldots, \mathbf{y}_{|\mathcal{Y}_0|-1}\} = \mathcal{Y}_0$, (It does not matter which element of $\mathcal{Y}_0$ is set to be e.g. $\mathbf{y}_0$ as long as the equality holds), and $\{\mathbf{y}_{|\mathcal{Y}_\ell|}, \ldots, \mathbf{y}_{|\mathcal{Y}_{\ell+1}|-1}\} = \mathcal{Y}_{\ell+1} - \mathcal{Y}_\ell$, $\ell \geq 1$.

We thus have our easily constructable quantizer codebook $\{\mathbf{y}_n\}$. We now show that this codebook can achieve the full-CSIT performance with finite-rate.

*D. Main results*

Let $\{\mathcal{F}_n\}$ be defined as in (3) and (4) with respect to $\{\mathbf{y}_n\}$. Also, let $\mathtt{b}_n^\star$ be as defined in Section III-B. We consider the quantizer $\mathtt{Q}^\star \triangleq \{\mathbf{y}_n, \mathcal{F}_n, \mathtt{b}_n^\star\}$.

**Theorem 1.** *For any $P > 0$, we have $\mathtt{OUT}(\mathtt{Q}^\star) = \mathtt{OUT}(\mathtt{Full})$ and $\mathtt{R}(\mathtt{Q}^\star) \leq \mathtt{C}_0(\frac{1}{P} + \frac{1}{P^t})$, where $\mathtt{C}_0$ is a constant that is independent of $P$.*

*Proof:* According to Proposition 1, and for any $\ell \in \mathbb{N}$, we have $\mathtt{OUT}(\mathtt{Q}^\star) = \mathtt{P}(\mathbf{h} \in \bigcap_{n\in\mathbb{N}} \mathcal{O}_{\mathbf{y}_n}) \leq \mathtt{P}(\mathbf{h} \in \bigcap_{\mathbf{y}\in\mathcal{Y}_\ell} \mathcal{O}_{\mathbf{y}})$, since by construction, $\mathcal{Y}_\ell \subset \{\mathbf{y}_n\}$. Now,

$$\mathtt{P}\left(\mathbf{h} \in \bigcap_{\mathbf{y}\in\mathcal{Y}_\ell} \mathcal{O}_{\mathbf{y}}\right) = \mathtt{P}(\forall \mathbf{y} \in \mathcal{Y}_\ell, |\langle\mathbf{y}, \overline{\mathbf{h}}\rangle|^2 \|\mathbf{h}\|^2 < \alpha) \leq \mathtt{P}\left(\|\mathbf{h}\|^2 < \frac{\alpha}{1 - 2^t/2^\ell}\right),$$

where $\overline{\mathbf{h}} = \mathbf{h}/\|\mathbf{h}\|$ and the last inequality follows from Proposition 2 for sufficiently large $\ell$. Since $\ell$ can be chosen arbitrarily large, we obtain $\mathtt{OUT}(\mathtt{Q}^\star) \leq \mathtt{P}(\|\mathbf{h}\|^2 < \alpha) = \mathtt{OUT}(\mathtt{Full})$. Since, $\mathtt{OUT}(\mathtt{Q}^\star) \geq \mathtt{OUT}(\mathtt{Full})$ is obvious, we have $\mathtt{OUT}(\mathtt{Q}^\star) = \mathtt{OUT}(\mathtt{Full})$.

We need to estimate the probabilities $P(\mathbf{h} \in \mathcal{F}_n)$ in order to evaluate $\mathtt{R}(\mathtt{Q}^\star)$. It is difficult to evaluate them individually, but as it turns out, we can estimate the "tail" sum-probability instead. In the appendix, we show that there is a constant $\ell_0 \geq 1$ such that for all $\ell \geq \ell_0$, we have $\sum_{n=|\mathcal{Y}_\ell|}^\infty P(\mathbf{h} \in \mathcal{F}_n) \leq \mathtt{C}_1 \alpha^t/2^\ell$, where $\mathtt{C}_1$ is a constant that is independent of $n$. For the remaining $1 \leq n \leq |\mathcal{Y}_{\ell_0}| - 1$, we use the trivial individual bounds $P(\mathbf{h} \in \mathcal{F}_n) \leq P(\mathbf{h} \in \mathcal{O}_{\mathbf{y}_0}) = 1 - e^{-\alpha} \leq \alpha$ that in fact holds for any $n \geq 1$, where the first inequality is by the definition in (4). These give us

$$\mathtt{R}(\mathtt{Q}^\star) = \sum_{n \in \mathbb{N}} P(\mathbf{h} \in \mathcal{F}_n) \mathtt{L}(\mathbf{b}_n^\star)$$

$$= \sum_{n=1}^\infty P(\mathbf{h} \in \mathcal{F}_n) \lfloor \log_2(n+1) \rfloor$$

$$= \sum_{n=1}^{|\mathcal{Y}_{\ell_0}|-1} P(\mathbf{h} \in \mathcal{F}_n) \lfloor \log_2(n+1) \rfloor + \sum_{\ell=\ell_0}^\infty \sum_{n=|\mathcal{Y}_\ell|}^{|\mathcal{Y}_{\ell+1}|-1} P(\mathbf{h} \in \mathcal{F}_n) \lfloor \log_2(n+1) \rfloor$$

$$\leq \lfloor \log_2 |\mathcal{Y}_{\ell_0}| \rfloor \sum_{n=1}^{|\mathcal{Y}_{\ell_0}|-1} P(\mathbf{h} \in \mathcal{F}_n) + \sum_{\ell=\ell_0}^\infty \lfloor \log_2 |\mathcal{Y}_{\ell+1}| \rfloor \sum_{n=|\mathcal{Y}_\ell|}^{|\mathcal{Y}_{\ell+1}|-1} P(\mathbf{h} \in \mathcal{F}_n)$$

$$\leq \alpha \lfloor \log_2 |\mathcal{Y}_{\ell_0}| \rfloor (|\mathcal{Y}_{\ell_0}| - 2) + \mathtt{C}_1 \alpha^t \sum_{\ell=\ell_0}^\infty \lfloor \log_2 |\mathcal{Y}_{\ell+1}| \rfloor / 2^\ell,$$

where for the first inequality, we have used the monotonicity of $\log_2(n+1)$, and for the second inequality we have applied the aforementioned tail sum-probability and trivial individual bounds regarding $\mathcal{F}_n$. The upper bound on $|\mathcal{Y}_\ell|$ in Proposition 2 implies that the sum in the last inequality is finite. This concludes the proof. ∎

It is difficult to calculate the minimum rate that guarantees the full-CSIT performance. However, bounds can be given that determines how this minimum rate behaves with $P$. Such bounds are useful as one is usually interested in the medium-to-high $P$ regime, where the outage probability is naturally low.

**Theorem 2.** *For any quantizer* $\mathtt{Q} = \{\mathbf{x}_n, \mathcal{E}_n, \mathbf{b}_n\}_\mathcal{I}$ *with* $\mathcal{I} \subset \mathbb{N}$ *and* $\mathtt{R}(\mathtt{Q}) \in o(\frac{1}{P})$, *we have* $\mathtt{OUT}(\mathtt{Q}) > \mathtt{OUT}(\mathtt{Full})$ *for sufficiently large* $P$.

*Proof:* Suppose that $\mathtt{R}(\mathtt{Q}) < 1$ (This holds by hypothesis for sufficiently large $P$). It is straightforward to show by contradiction that there is an index $i \in \mathcal{I}$ such that $P(\mathbf{h} \in \mathcal{E}_i) \geq 1 - \mathtt{R}(\mathtt{Q})$. Without loss of generality, suppose that $P(\mathbf{h} \in \mathcal{E}_1) \geq 1 - \mathtt{R}(\mathtt{Q})$ and $1 \in \mathcal{I}$. Then, with $f(\mathbf{h}) = \frac{1}{\pi^t} e^{-\|\mathbf{h}\|^2}$ representing the PDF of $\mathbf{h}$, we have

$$\mathtt{OUT}(\mathtt{Q}) = \sum_{i \in \mathcal{I}} \int_{\mathcal{E}_i} \mathbf{1}\left(|\langle \mathbf{x}_i, \mathbf{h} \rangle|^2 < \alpha\right) f(\mathbf{h}) d\mathbf{h}$$

$$\geq \int_{\mathcal{E}_1} \mathbf{1}\left(|\langle \mathbf{x}_1, \mathbf{h} \rangle|^2 < \alpha\right) f(\mathbf{h}) d\mathbf{h}$$

$$= \int_{\mathbb{C}^T} \mathbf{1}\left(|\langle \mathbf{x}_1, \mathbf{h} \rangle|^2 < \alpha\right) f(\mathbf{h}) d\mathbf{h} - \int_{\mathcal{E}_1^c} \mathbf{1}\left(|\langle \mathbf{x}_1, \mathbf{h} \rangle|^2 < \alpha\right) f(\mathbf{h}) d\mathbf{h}$$

$$\geq \mathtt{OUT}(\mathtt{Open}) - \int_{\mathcal{E}_1^c} f(\mathbf{h}) d\mathbf{h}$$

$$\geq \mathtt{OUT}(\mathtt{Open}) - \mathtt{R}(\mathtt{Q}),$$

Since $\mathtt{OUT}(\mathtt{Open}) \in \Theta(\frac{1}{P})$ and $\mathtt{R}(\mathtt{Q}) \in o(\frac{1}{P})$, the final lower bound is $\Theta(\frac{1}{P})$, while it is well-known that $\mathtt{OUT}(\mathtt{Full}) \in \Theta(\frac{1}{P^t})$. This concludes the proof. ∎

By Theorems 1 and 2 we thus conclude that the necessary and sufficient feedback rate to achieve the full-CSIT performance is $\Theta(\frac{1}{P})$.

## IV. Conclusions and Extensions

We have identified a practical vector quantizer design problem where any fixed-length quantizer (FLQ) yields non-zero distortion at any finite rate, while there is a variable-length quantizer (VLQ) that can achieve zero distortion with arbitrarily low rate: In a point-to-point $t \times 1$ multiple antenna system with quantized channel state information at the transmitter (CSIT), we have shown that there is a VLQ that achieves the full-CSIT outage probability performance with rate $O(\frac{1}{P})$. We have also shown that such decay of the feedback rate is $P$-asymptotically the best possible.

An interesting problem is to determine the minimum feedback rate where the full-CSIT performance is possible. There appears to be a lot of room for improvement in this context. Indeed, using random vector quantizers, we can show that almost all infinite-cardinality codebooks with the encoding strategy presented in this paper can achieve the full-CSIT performance (the "bad" codebooks have zero measure). Due to space limitations, we shall discuss this issue in the journal version of this paper.

Another extension is to investigate the performance with the more general precoding strategy. In precoding, one uses a $t \times t$ precoding matrix $\mathbf{X}$ with $\|\mathbf{X}\| = 1$ instead of a beamforming vector. An outage event occurs if $\|\mathbf{X}\mathbf{h}\|^2 < \alpha$. Using the beamforming vector $\mathbf{x}$ is thus equivalent to using the precoding matrix $\mathbf{x}\mathbf{x}^\dagger$. In such a scenario, it can be shown that the necessary and sufficient feedback rate that guarantees the full-CSIT performance is $\Theta(\frac{1}{P^t})$ instead. The achievability is the same as Theorem 1 with the first beamforming vector $\mathbf{y}_0$ in the quantizer codebook replaced by the identity precoding matrix. The converse follows the exact same arguments that are used in the proof of Theorem 2.

Finally, note that we have allowed the use of non-prefix codes for the feedback binary codewords. These codes are well-suited for channel quantization purposes as one does not concatenate codewords for different channel states. On the other hand, the applicability of an empty codeword appears to be a gray area in practical systems. Our results can however be extended to the case where empty codeword is not allowed or even the case where only prefix-free codes are to be used. For example, for Theorem 1, instead of the non-prefix-free code, one can use the prefix-free code discussed in Example 2. Then, $\mathtt{Q}^\star$ will have a feedback rate of $1 + O(\frac{1}{P})$ instead of $O(\frac{1}{P})$. An extension of Theorem 2 then reveals that the $1 + \Theta(\frac{1}{P})$ feedback rate is in fact asymptotically the best possible.

## Appendix

Here, we estimate the tail sum-probabilities corresponding to the sequence $\{\mathcal{F}_n\}$. Consider an arbitrary vector $\mathbf{h}_0 \in \mathbb{C}^t$ with $\|\mathbf{h}_0\|^2 > \alpha(1 + \frac{3t}{2^\ell})$. According to Proposition 2, for sufficiently large $\ell$, there is a vector $\mathbf{y} \in \mathcal{Y}_\ell$ with

$$|\langle \mathbf{y}, \mathbf{h}_0 \rangle|^2 > \|\mathbf{h}_0\|^2 \left(1 - \frac{2t}{2^\ell}\right).$$

Using the fact that $\|\mathbf{h}_0\|^2 > \alpha(1 + \frac{3t}{2^\ell})$, we have, for sufficiently large $\ell$,

$$|\langle \mathbf{y}, \mathbf{h}_0\rangle|^2 > \alpha\left(1 + \frac{3t}{2^\ell}\right)\left(1 - \frac{2t}{2^\ell}\right)$$
$$= \alpha\left(1 + \frac{t}{2^\ell} - \frac{6t^2}{2^{2\ell}}\right)$$
$$> \alpha,$$

which implies $\mathbf{h}_0 \in \mathcal{O}_{\mathbf{y}}^c$. In other words, for any $\mathbf{h}$ with $\|\mathbf{h}_0\|^2 > \alpha(1+\frac{3t}{2^\ell})$, there exists $\mathbf{y} \in \mathcal{Y}_\ell$ such that $\mathbf{h} \in \mathcal{O}_{\mathbf{y}}^c$. Therefore,

$$\left\{\mathbf{h} \in \mathbb{C}^t : \|\mathbf{h}\|^2 > \alpha\left(1 + \frac{3t}{2^\ell}\right)\right\} \subset \bigcup_{\mathbf{y} \in \mathcal{Y}_\ell} \mathcal{O}_{\mathbf{y}}^c. \tag{12}$$

On the other hand,

$$\{\mathbf{h} \in \mathbb{C}^t : \|\mathbf{h}\|^2 < \alpha\} \subset \mathcal{O}_{\mathbf{y}}, \ \forall \mathbf{y} \in \mathcal{X}_{\mathsf{BF}},$$

and therefore,

$$\{\mathbf{h} \in \mathbb{C}^t : \|\mathbf{h}\|^2 < \alpha\} \subset \bigcap_{i \in \mathbb{N}} \mathcal{O}_{\mathbf{y}_i}. \tag{13}$$

Now,

$$\bigcup_{i=0}^{|\mathcal{Y}_\ell|-1} \mathcal{F}_i = \bigcap_{i\in\mathbb{N}} \mathcal{O}_{\mathbf{y}_i} \cup \bigcup_{i=0}^{|\mathcal{Y}_\ell|-1} \mathcal{O}_{\mathbf{y}_i}^c$$
$$\supset \left\{\mathbf{h} \in \mathbb{C}^t : \|\mathbf{h}\|^2 < \alpha \text{ or } \|\mathbf{h}\|^2 > \alpha\left(1 + \frac{3t}{2^\ell}\right)\right\},$$

where the equality is by the definition of $\{\mathcal{F}_n\}$ in (3) and (4), and the last inclusion follows from (12) and (13). This implies

$$\bigcup_{i=|\mathcal{Y}_\ell|}^{\infty} \mathcal{F}_i = \left(\bigcup_{i=0}^{|\mathcal{Y}_\ell|-1} \mathcal{F}_i\right)^c \subset \left\{\mathbf{h} \in \mathbb{C}^t : \alpha \leq \|\mathbf{h}\|^2 \leq \alpha\left(1 + \frac{3t}{2^\ell}\right)\right\}.$$

Therefore,

$$\mathsf{P}\left(\mathbf{h} \in \bigcup_{i=|\mathcal{Y}_\ell|}^{\infty} \mathcal{F}_i\right) \leq \int_\alpha^{\alpha(1+\frac{3t}{2^\ell})} \frac{x^{t-1}e^{-x}}{\Gamma(t)}\mathrm{d}x$$
$$< \int_\alpha^{\alpha(1+\frac{3t}{2^\ell})} x^{t-1}\mathrm{d}x$$
$$= \frac{\alpha^t}{t}\left[\left(1 + \frac{3t}{2^\ell}\right)^t - 1\right]$$
$$= \frac{3\alpha^t}{t2^\ell}\underbrace{\left[1 + \left(1 + \frac{3t}{2^\ell}\right) + \cdots + \left(1 + \frac{3t}{2^\ell}\right)^{t-1}\right]}_{<t+1 \text{ for sufficiently large } \ell}$$
$$< \mathtt{C}_1 \frac{\alpha^t}{2^\ell},$$

where the last inequality holds for sufficiently large $\ell$, and $\mathtt{C}_1$ is a constant that is independent of $\ell$.

## References

[1] R. M. Gray and D. L. Neuhoff, "Quantization," *IEEE Trans. Inf. Theory*, vol. 44, no. 6, pp. 2325–2383, June 1998.

[2] V. Misra, V. K. Goyal, and L. R. Varshney, "Distributed scalar quantization for computing: High-resolution analysis and extensions," *IEEE Trans. Inf. Theory*, vol. 57, no. 8, pp. 5298–5325, Aug. 2011.

[3] D. J. Love, R. W. Heath, Jr., V. K. N. Lau, D. Gesbert, B. D. Rao, and M. Andrews, "An overview of limited feedback in wireless communication systems," *IEEE J. Select. Areas Commun.*, vol. 26, no. 8, pp. 1341–1365, Oct. 2008.

[4] D. J. Love, R. W. Heath, Jr., and T. Strohmer, "Grassmannian beamforming for multiple-input multiple-output wireless systems," *IEEE Trans. Inf. Theory*, vol. 49, no. 10, pp. 2735–2747, Oct. 2003.

[5] S. Zhou, Z. Wang, and G. B. Giannakis, "Quantifying the power loss when transmit beamforming relies on finite-rate feedback," *IEEE Trans. Wireless Commun.*, vol. 4, no. 4, pp. 1948–1957, Jul. 2005.

[6] A. Narula, M. J. Lopez, M. D. Trott, and G. W. Wornell, "Efficient use of side information in multiple antenna data transmission over fading channels," *IEEE J. Select. Areas Commun.*, vol. 16, no. 8, pp. 1423–1436, Oct. 1998.

[7] J. C. Roh and B. D. Rao, "Transmit beamforming in multiple antenna systems with finite rate feedback: A VQ-based approach," *IEEE Trans. Inf. Theory*, vol. 52, no. 3, pp. 1101–1112, Mar. 2006.

[8] V. K. N. Lau, Y. Liu, and T.-A. Chen, "On the design of MIMO block-fading channels with feedback-link capacity constraint," *IEEE Trans. Commun.*, vol. 52, no. 1, pp. 62–70, Jan. 2004.

[9] J. Zheng, E. R. Duni, and B. D. Rao, "Analysis of multiple-antenna systems with finite-rate feedback using high-resolution quantization theory," *IEEE Trans. Signal Process.*, vol. 55, no. 4, pp. 1461–1476, Apr. 2007.

[10] C. K. Au-Yeung and D. J. Love, "On the performance of random vector quantization limited feedback beamforming in a MISO system," *IEEE Trans. Wireless Commun.*, vol. 6, no. 2, pp. 458–462, Feb. 2007.

[11] K. K. Mukkavilli, A. Sabharwal, and E. Erkip, "On beamforming with finite-rate feedback for multiple antenna systems," *IEEE Trans. Inf. Theory*, vol. 49, no. 10, pp. 2562–2579, Oct. 2003.

[12] P. Xia and G. B. Giannakis, "Design and analysis of transmit-beamforming based on limited-rate feedback," *IEEE Trans. Signal Process.*, vol. 54, no. 5, pp. 1853–1863, May 2006.

[13] S. A. Jafar and S. Srinivasa, "On the optimality of beamforming with quantized feedback," *IEEE Trans. Commun.*, nol. 55, no. 12, pp. 2288–2302, Dec. 2007.